\documentclass[twocolumn,showpacs,preprintnumbers,amsmath,amssymb]{revtex4}
\usepackage{graphicx}
\usepackage{dcolumn}
\usepackage{bm}
\usepackage{comment}
\usepackage{subfigure}
\usepackage{amsmath}
\usepackage{url}
\usepackage{hyperref}
\usepackage{breakurl}

\begin{document}

\title{The effect of guide-field and boundary conditions on collisionless magnetic reconnection in a stressed $X$-point collapse} 
\author{J. Graf von der Pahlen and D. Tsiklauri}
\affiliation{School of Physics and Astronomy, Queen Mary University of London, London, E1 4NS, United Kingdom}
\date{\today}
\begin{abstract}
Works of D. Tsiklauri, T. Haruki, Phys. of Plasmas, 15, 102902 (2008) and D. Tsiklauri and T. Haruki, Phys. of Plasmas, 14, 112905, (2007) are extended by inclusion of the out-of-plane magnetic (guide) field. In particular, magnetic reconnection during collisionless, stressed $X$-point collapse for varying out-of-plane guide-fields is studied using a kinetic, 2.5D, fully electromagnetic, relativistic particle-in-cell numerical code. For zero guide-field, cases for both open and closed boundary conditions are investigated, where magnetic flux and particles are lost and conserved respectively. It is found that reconnection rates, out-of-plane currents and density in the $X$-point increase more rapidly and peak sooner in the closed boundary case, but higher values are reached in the open boundary case. The normalized reconnection rate is fast: 0.10-0.25. In the open boundary case it is shown that an increase of guide-field yields later onsets in the reconnection peak rates, while in the closed boundary case initial peak rates occur sooner but are suppressed. The reconnection current changes similarly with increasing guide-field, however for low guide-fields the reconnection current increases, giving an optimal value for the guide-field between 0.1 and 0.2 times the in-plane field in both cases. Also, in the open boundary case, it is found that for guide-fields of the order of the in-plane magnetic field, the generation of electron vortices occurs. Possible causes of the vortex generation, based on the flow of decoupled particles in the diffusion region and localized plasma heating, are discussed. Before peak reconnection onset, oscillations in the out-of-plane electric field at the $X$-point are found, ranging in frequency from approximately 1 to 2 $\omega_{pe}$ and coinciding with oscillatory reconnection. These oscillations are found to be part of a larger wave pattern in the simulation domain. Mapping the out-of-plane electric field along the central lines of the domain over time and applying a 2D Fourier transform reveals that the waves predominantly correspond to the ordinary and the extraordinary mode and hence may correspond to observable radio waves such as solar radio burst fine structure spikes. 
\end{abstract}


\maketitle

\section{Introduction}

\pagestyle{plain}
\thispagestyle{plain}

Dungey's original work on X-type collapse \cite{Dungey53} is the earliest analysis of magnetic reconnection, predating tearing-mode and Sweet-Parker theories, and X-type scenarios are still frequently considered in models of solar flares (see Ref. \cite{aschwanden}, chapter 10.5.1). Most space and solar plasma simulations of magnetic reconnection predominantly focus on reconnection induced by the tearing-mode instability. Yet, after a Harris type current sheet is disrupted by a tearing instability and magnetic islands and $X$-points start to form, there are indeed few distinguishable differences between $X$-point collapse and the well-studied tearing instability. In both cases a stage is reached where $X$-point symmetry is broken, which means that there is no restoring force and the $X$-point collapses, resulting in fast reconnection (see Ref. \cite{priest}, chapter 7.1). Moreover, even the respective causes of the reconnection electric field, as calculated using the generalized Ohm's law, are the same, namely the off-diagonal terms of the divergence of the electron pressure tensor. This was shown in Ref. \cite{tsiklauri2008} for the case of $X$-point collapse and in Ref. \cite{Prittchet} for the tearing-mode instability. Equally similar is the quadrupolar structure of the out-of-plane magnetic field at the $X$-point, caused by the Hall effect (first proposed in Ref. \cite{sonnerup} and shown to be critical in fast reconnection in Ref. \cite{mandt}). This was shown to emerge in the case of $X$-point collapse in Ref. \cite{Tsiklauri2007} and in the case of tearing-mode instability in Ref. \cite{Prittchet}, as well as in the Magnetic Reconnection Experiment (MRX) \cite{tharp2}.

$X$-point collapse is distinguishable as a reconnection scenario by its geometry, as illustrated in Ref. \cite{priest}, chapter 2.1. In a $X$-point collapse set-up, field lines are of the shape of contracted rectangular hyperbolas, with the field strength decreasing towards the origin i.e. the $X$-point. This set-up, for a two-dimensional grid centred on the origin, can be defined most simply as  

    \begin{equation}
    B_x = y, B_y = \alpha^2x, \;\;\; 
    \label{eqn:bxcol}
    \end{equation}
with a corresponding out-of-plane current, based on Ampere's law, of
    \begin{equation}
    j_z = \frac{\alpha^2-1}{\mu}. \;\;\; 
    \label{eqn:jzxcol}
    \end{equation}   
Here, $\alpha$ being greater or smaller than 1 corresponds to a contraction along the $x$ or $y$ axis respectively, and results in an inwards $j\times B$ force on the plasma along the same axis. Field lines are dragged along by the plasma, further increasing the contraction. With the resulting build-up of magnetic pressure near the $X$-point, magnetic reconnection commences and the field collapses. 


This study extends the works of D. Tsiklauri  and T. Haruki (\cite{tsiklauri2008} and \cite{Tsiklauri2007}), where $X$-point collapse was modelled using a fully relativistic 2.5D Particle in Cell (PIC) code, by introducing an uniform out-of-plane magnetic guide-field of varying magnitudes and by comparing the effects of different boundary conditions (details of the code and set-up used are given in \ref{sec:model}).

\begin{figure}[htbp]
    \includegraphics[width=0.99\linewidth]{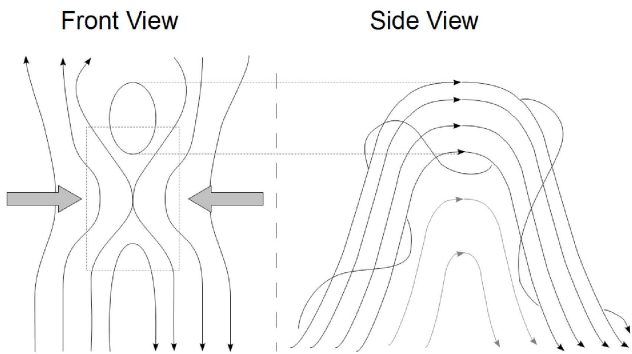}
     \caption{ \label{Fig:prominence}Adopted from \cite{aschwanden}, p. 437, based on \cite{hirayama74}, showing a snap shot of Hirayama's model of solar flares, indicating reconnection site (front view) and guide-field (side view).} 
      \end{figure}

The relevance of reconnection with a guide-field in natural reconnection processes is apparent in solar flare models such as that of Hirayama \cite{hirayama74}, as shown in Fig.~\ref{Fig:prominence}. In this model a rising solar prominence stretches out a current sheet, prone to trigger magnetic reconnection (see Ref. \cite{aschwanden}, chapter 10.5.1). Here it is clear that the filaments making up the prominence are mostly perpendicular to the plane of the reconnection site (or rather parallel to the current sheet) and thus constitute a guide-field. Similarly guide-fields play a role in reconnection in the magnetotail of the Earth, as demonstrated in Ref. \cite{grigorenko}. 

The need to account for kinetic effects in a simulation of $X$-point collapse is explained in Ref. \cite{priest}, chapter 7.1.1. As shown here, it was demonstrated in Ref. \cite{craig} that the condition
    \begin{equation} 
    \beta_0 \lesssim \left(\frac{\eta}{v_{A0}y_0}\right)^{0.565},
    \label{eqn:condition}
    \end{equation}
where usual notation is used, must apply for a sufficiently thin current sheet to form, allowing for fast-reconnection to be possible in a $X$-point collapse scenario. If this condition is not satisfied, the collapse of the $X$-point will be counteracted by the plasma pressure and choked off. In the case of solar plasmas the right hand side of this condition is of the order of $10^{-8}$, implying that $\beta \lesssim 10^{-8}$, which is much smaller than values based on observation. Hence, kinetic effects play a decisive role if reconnection is to be fast \cite{Tsiklauri2007,tsiklauri2008}. Further, the importance of using a kinetic simulation is apparent from the shortcomings of purely MHD based simulations, as described in Ref. \cite{birn} chapter 3.1.1. In particular, MHD models tend to have dissipation regions of small width to length ratios (e.g. Sweet-Parker), which do not allow for fast reconnection. On the other hand, when they have suitable diffusion region width to length ratios (Petcheck), they rely on non-uniform resistivity.


%

There have been several other studies into the matter of magnetic reconnection in the case of a guide-field, with different set-ups, including computational studies by Horiuchi et al. \cite{sato97}, Pritchett et al. \cite{pritchett.guidefield,pritchett.vortex} and Fermo et al. \cite{fermo}, as well as experimental studies using the MRX \cite{tharp2,Tharp2012} and the Versatile Toroidal Facility (VTF) \cite{vtf,vtf.langmuir}. While the field configurations in these studies were distinctly different from those presented here, some of their findings are similar, including the delays in onset times of peak reconnection \cite{sato97,Tharp2012} and the increase in out-of-plane current along one pair of separatrix arms \cite{sato97,pritchett.guidefield} (see section \ref{sec:mainresults}). Further, electron vortices were shown to emerge at the $X$-point for guide fields approximately equal to the in-plane field, similar to vortex formation simulated in Ref. \cite{fermo} (see Ref. \ref{sec:vortex}). These vortices appear singularly at the $X$-point, as opposed to in multiples along the $X$-line, and thus appear distinctly different from vortices observed in Ref. \cite{pritchett.vortex}. While (as shown in Ref. \cite{fermo} and \cite{pritchett.vortex}) vortices were found previously in similar numerical set-ups, we find for the first time that imposing a guide-field can also lead to electron vortex formation in X-point Collapse. Albeit, this also underscores the similarities between $X$-point collapse and tearing-mode set-ups and that more attention needs to be paid to the former.

As discussed in Ref. \cite{birn} page 91-94, the generation of electromagnetic waves plays a significant role in Hall reconnection scenarios. The generation of whistler and Alfv\'en waves during reconnection was simulated in Ref. \cite{whistleralven} and demonstrated experimentally at the VTF \cite{vtf}. Further, electro-static waves were detected at the $X$-point in a VTF experiment during reconnection in Ref. \cite{vtf.langmuir}. As shown in section \ref{sec:oscil}, both electrostatic and electromagnetic waves are produced directly preceding reconnection onset, with the electromagnetic waves being in the radio frequency regime. These could be similar to radiation from events such as Type \textrm{II} precursors (first discovered in Ref. \cite{precursor}), which are known to release radio waves prior to solar flares and thus prior to reconnection in the Hirayama model.


\section{Simulation model}
  \label{sec:model}

\subsection{Stressed $X$-point Collapse Reconnection Model}
 \label{sec:xcol}

  \begin{figure}[htbp]
    \includegraphics[width=0.9\linewidth]{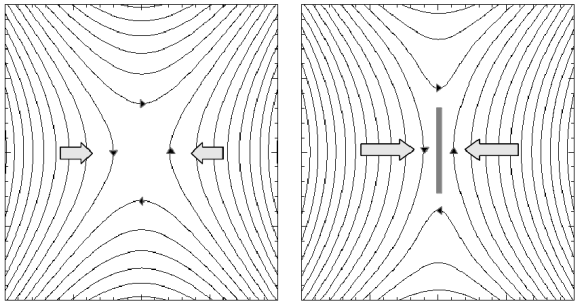}
     \caption{ \label{Fig:fieldcollapse}Magnetic field line configuration at the start of the simulation (left) and at the peak of the reconnection (right). Directions of field lines are indicated. Arrows indicate the velocity of the field lines due to the $\bf{J}\times\bf{B}$ force. The grey vertical strip indicates the current sheet.} 
   \end{figure}
  
The left panel of Fig.~\ref{Fig:fieldcollapse} shows the initial set-up of the in-plane magnetic field used in this study. This is mathematically described by the following expressions
    \begin{equation}
    B_x = \frac{B_0}{L} y,
    \;\;\;    B_y = \frac{B_0}{L} \alpha^2 x, \;\;\; 
    \label{eqn:Bxy}
    \end{equation}
where $B_0$ is magnetic field intensity at the distance $L$ from the $X$-point for $\alpha = 1.0$, $L$ is the global external length-scale of reconnection, and $\alpha$ is the stress parameter (see e.g. chapter 2.1 in Ref. \cite{birn}). In addition a uniform current is imposed at time $t=0$ in the $z$-direction, corresponding to the curl of the magnetic field, such that Ampere's law is satisfied

  \begin{equation}
    j_z = \frac{B_0}{\mu_0 L} (\alpha^2 - 1).
   \end{equation}

The configuration above was simulated and analysed in Ref. \cite{Tsiklauri2007}. In this scenario the initial stress in the field leads to a $\bf{J}\times\bf{B}$ force that pushes the field lines horizontally inwards. This serves to increase the initial imbalance, which in turn increases the inwards force and the field collapses. Due to the frozen-in condition, this leads to a build up of plasma near the $X$-point and eventually to the formation of a diffusion region, accompanied by a current sheet as shown in the right panel of Fig.~\ref{Fig:fieldcollapse}. The term in the generalized Ohm's law which corresponds to the breaking mechanism of the frozen in condition was shown to be the off-diagonal terms of the electron pressure tensor divergence, due to electron meandering motion (see Ref. \cite{tsiklauri2008}).  

The new effect introduced in this study is an out-of-plane magnetic guide-field of different intensities. The strengths of the guide-field were chosen to be fractions of the maximum field amplitude within the plane, $B_{P}$, i.e.     

    \begin{equation} 
     B_{Z0} = (n/10)B_0\sqrt{1+\alpha^2}=(n/10)B_{P},
    \label{eqn:Bz}
    \end{equation}
where $n$ is an integer ranging from 1 to 10.

  \subsection{PIC Simulation Code}
  \label{sec:pic}
  
  The simulation code used here is a 2.5D relativistic and fully electromagnetic Particle In Cell (PIC) code, as developed by the EPOCH collaboration, based on the original PSC code by Hartmut Ruhl \cite{ruhl2006} (minor modifications were made in this study to allow for closed boundary conditions). Following the general workings of a PIC code, the particles making up the plasma are simulated as a smaller number of 'quasi-particles', representing the distribution function of position and momentum of a given particle species. The electric and magnetic fields are defined on a grid with a size according to the simulation domain and a resolution determined by the size of the grid cells. The quasi-particles are moved according to the self-consistent electric and magnetic fields, defined using a finite difference time domain technique. From the particle momenta, currents are calculated using the Villasenor and Buneman scheme \cite{Bunemann1992}, which are then used to update the electric and magnetic fields at the end of the time step.   

The number density of both electrons and ions in the simulation domain, $n_e$ and $n_p$, was set to $10^{16}m^{-2}$, while the temperature for both electrons and protons, $T_e$ and $T_p$, was set to $6.0\times10^7$K, matching conditions of flaring in the solar corona. The proton mass was set to 100 times the electron mass, i.e. $m_p = 100m_e$, to speed up the code. The value of $B_0$ was set as $0.03207$ T such that $\omega_{pe} = \omega_{ce}$, where $\omega_{ce}$ is the electron cyclotron frequency, which fixes the Alfv\'en speed at the $y$-boundary as $v_a = 0.1c$. The size of the grid cells in both $x$ and $y$ where set to the Debye length, $\lambda_D$, such that

    \begin{equation} 
    dx=dy=\lambda_D = v_{te}/\omega_{pe}=\sqrt{k_BT_e\epsilon_0/n_ee^2}
    \label{eqn:gridsize}
    \end{equation}
where usual notation is used. The timestep in this simulation was determined by the simulation code and set as

\begin{equation}
    dt = \frac{\lambda_D}{c\sqrt{2}},
    \end{equation}
where $c$ is the speed of light in vacuum. 

The simulation used a grid of 400 $\times$ 400 cells, which thus gave the simulation domain a length of $400\times\lambda_D$, according to ($\ref{eqn:gridsize}$), i.e. $2L=2.1382$ m, corresponding to 4.0208$c/\omega_{pi}$. The code used 500 particles per species per cell and thus a total $1.6\times10^8$ particles in total. While the simulation in Ref. \cite{Tsiklauri2007} used only 100 particles per cell, which is sufficient to accurately resolve electromagnetic field dynamics, it was found that a greater number was needed to accurately resolve particle dynamics, particularly with regards to the discovery of the electron vortex (see section \ref{sec:vortex}). Convergence tests showed that results converged with those at 1000 particles per cell.

 \subsection{Boundary Conditions}
 \label{sec:bc}

The original simulation of Tsiklauri and Haruki (see Ref. \cite{Tsiklauri2007}) used boundary conditions such that flux at the boundary is conserved. Hence, zero-gradient boundary conditions are imposed 
  both on the electric and magnetic fields 
  in  $x$- and $y$-directions and the tangential component of electric field 
  was forced to zero, while the normal component of the magnetic field was kept
  constant. This allows for magnetic field lines to slide freely on the boundaries, whilst inhibiting loss (or gain) of magnetic flux from the simulation domain. Further, the boundary condition for (quasi) particles in the simulation was set so that particles are reflected when reaching the boundary. Thus, the the simulation represents an isolated physical system, neither losing nor gaining magnetic or particle flux (we will refer to this as closed boundary conditions). 

We also consider boundary conditions that allow outflow of flux and particles i.e. where waves or particles travelling past the simulation boundary are simply removed and no longer affect the simulation (we will refer to this as open boundary conditions). Both this and the previous case were investigated, with and without guide-field, such that differences could be established. 

Based on simple physical considerations it can be argued that open boundary conditions are more relevant in many astrophysical reconnection scenarios, as for example in the solar flare reconnection model described in Ref. \cite{hirayama74}, or more generally the space around tearing-unstable $X$-points, where there is no physical restriction on the motion of field lines and hence particles. 

On the other hand, closed boundary condition are more relevant to reconnection in confined reconnection experiments, particularly 'closed-type' experiments (tokamaks, spheromaks etc.) where motion of field lines and particles is restricted. Many of these experimental set-ups and the effects of their specific boundary conditions are discussed in Ref. \cite{ono}.

\section{Simulation Results}
\label{sec:results}

 \subsection{Effect of Boundary Conditions}
 \label{sec:bceffect}

\begin{figure}[htbp]
    \includegraphics[width=0.9\linewidth]{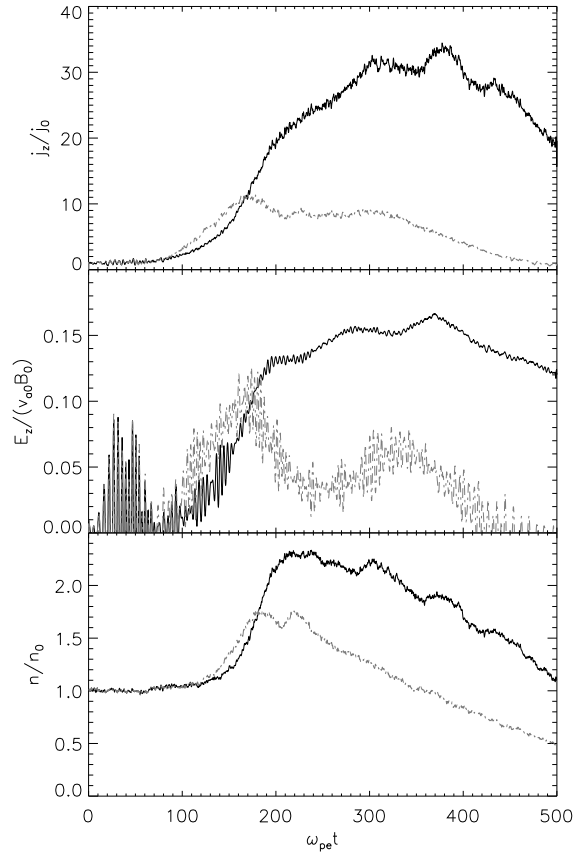}
     \caption{ \label{Fig:picbccompare}(top) and (middle) respectively show the reconnection electric field and current at the $X$-point as a function of time. In black the case for open boundaries is shown, whereas the grey line shows the case for closed boundaries. (Bottom) The particle number density at the $X$-point, also for both boundary cases. Again, black represents the open boundary case whereas grey represents the closed boundary case.}  
   \end{figure}

An initial test was carried out for the zero guide-field case, i.e. $B_{Z0}=0$, showing how the simulation results of the original study \cite{Tsiklauri2007} vary when applying open instead of closed boundary conditions, as shown in Fig.~\ref{Fig:picbccompare}. Reconnection electric field, reconnection current and density at the $X$-point appear to increase more rapidly and peak sooner in the closed boundary case. However, higher values appear to be reached in the open boundary for all three variables, particularly in the reconnection current. Also in the open boundary case, there is a much lesser decline within the simulation time and the characteristic double peak profile in the reconnection electric field in the closed boundary case is not apparent, at least not before $500\omega_{pe}^{-1}$. All of the above observed differences indicate that the change in boundary condition has a significant effect on the reconnection dynamics. By plotting the magnetic flux function (calculated using the spatial gradients of the magnetic field components and components of $E_Z$) as contours of equal magnitude, at different times in the simulation, these differences in reconnection were shown dynamically (see movie 1 and 2 in \cite{sup}). 

In both cases, rapid high-amplitude oscillations in the out-of-plane electric field occur immediately before reconnection onset. Plotting contours of the out-of-plane flux-function at the $X$-point, it was shown that this oscillation corresponds to oscillatory reconnection (see movie 3 in supplementary material \cite{sup}), as first demonstrated in Ref. \cite{oscilrecon2}. Further oscillations, with a smaller amplitude, emerge later in the simulation in both cases, but predominantly in the closed case, as shown in the middle panel of Fig.~\ref{Fig:picbccompare}. As opposed to reconnection in a uni-directionally sheared magnetic field, field lines in $X$-point collapse reaching the $X$-point are likely to carry different particle distributions due to their differences in initial geometry, which is a likely factor in the transient nature of the reconnection observed. In the open case, field lines are free to shift on the boundary (see movie 1 in \cite{sup}), which may reduce the impact of the geometry, thus reducing the oscillations in the reconnection electric field as shown (these oscillations and the generation of electromagnetic waves preceding reconnection onset will be discussed further in section \ref{sec:oscil}).

 \subsection{Effect of Guide-Field}

 \subsubsection{Effects on Reconnection Rate and shape of Current sheet}
 \label{sec:mainresults}

\begin{figure}[htbp]
        \includegraphics[width=0.9\linewidth]{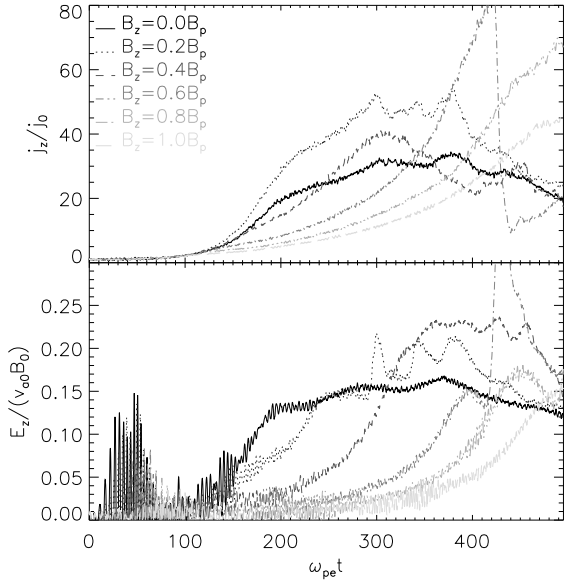} 
\caption{The reconnection current and electric field over time, for different levels of guide-field as indicated, for open boundary conditions.}%
\label{Fig:picezjz}%
\end{figure}

\begin{figure}[htbp]
        \includegraphics[width=0.9\linewidth]{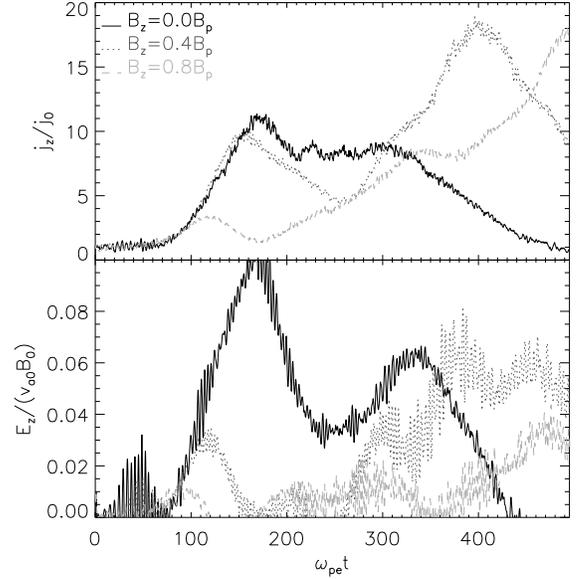} 
\caption{As in Fig.~\ref{Fig:picezjz} but for the closed boundary case.}%
\label{Fig:picezjzbc}%
\end{figure}

 The obtained values for reconnection electric field and current for the different strengths of guide-fields as a function of time are shown in Fig.~\ref{Fig:picezjz} for the open boundary case and in Fig.~\ref{Fig:picezjzbc} for the closed boundary case. In the open boundary case we see a trend of delayed on-set times in the reconnection field. The same is true for the reconnection current, except for $n=2$ (i.e. $B_{Z0}=0.2B_{P}$). In the closed boundary case initial peaks are reached sooner for greater values of guide-field, but their amplitudes are significantly reduced. Onset times of the secondary peak however are also delayed. 
 
 These results are consistent with other reconnection simulations using different set ups, including the 1997 study by Horiuchi and Sato \cite{sato97}. Here it is argued that the increase in guide-field reduces the orbit associated with the meandering motion of electrons, making them more magnetized and less likely to break the frozen-in condition and thus delaying on-set times. Further, the reduction in reconnection electric field in the closed boundary case is in-line with the findings in Ref. \cite{tharp2}, where it is shown that, in the MRX set-up, the reconnection electric field corresponds to that of a Hall MHD simulation and is reduced for greater guide-fields due to a reduction in the Hall-current.

At later stages in the open boundary case, the reconnection field for guide-fields of $B_{Z0} = 0.2$ up to $B_{Z0} = 0.6B_{P}$ tends to significantly exceed that of the zero-guide-field case. Further, we see somewhat chaotic peaks occurring, leading to an exceptionally high peak value at $B_{Z0} = 0.6B_{P}$. We discuss in section \ref{sec:vortex} how this particular peak, as well as the peak values for $B_{Z0} = 0.8$ and $B_{Z0} = 1.0B_{P}$, are linked to the emergence of an electron vortex at the $X$-point along with a magnetic island. This is of relevance since it has been shown in Ref. \cite{island.current} that magnetic islands, formed at the $X$-point, can have an effect on electron acceleration and lead to an increased out-of-plane current (see specifically the central island in Ref. \cite{island.current}, fig 6(c)).

The increase in reconnection current at $n=2$  was investigated further by running the simulation with guide-fields ranging from $0$ to $0.5B_P$ in steps of $0.05B_P$. It was shown that, for the open boundary case, the initial rate of increase of the reconnection current reached a peak value, giving an "optimal" guide-field strength between $0.1B_P$ and $0.2B_P$. An increase of the reconnection current caused by the introduction of a guide-field, as well as an increase in the current along one pair of separatrix arms, is also observed and discussed in detail in Ref. \cite{pritchett.guidefield}. It is discussed here how the diversion of particle flows due to the guide-field can result in the confinement of electrons to one of the separatrices, leading to a greater density and thus a greater out-of-plane current.

  \begin{figure}[htbp]
    \includegraphics[width=0.99\linewidth]{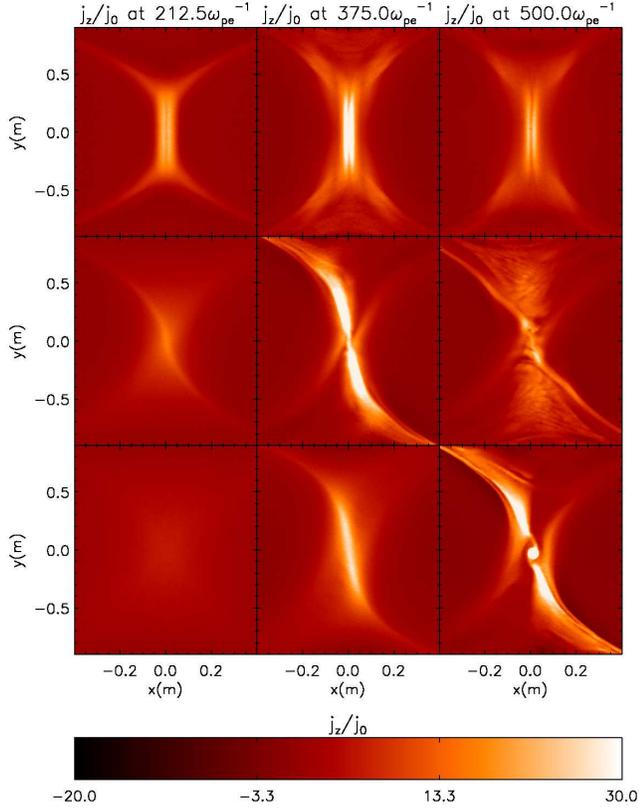}
     \caption{ \label{Fig:jzcon}Out-of-plane current in the open boundary case for different guide-field strengths at different simulation times. The three columns correspond from left to right to simulation times of $212.5\omega_{pe}^{-1}$, $375.0\omega_{pe}^{-1}$ and $500.0\omega_{pe}^{-1}$ respectively. The three rows correspond from top to bottom to guide field strengths of $0.0$, $0.4B_{P}$ and $0.8B_{P}$ respectively.} 
      \end{figure}

Fig.~\ref{Fig:jzcon} shows the out-of-plane current in the simulation domain for the open boundary case for three different strengths of guide-field at three different snapshots in time. The plots show that for greater guide-fields onset times are delayed and distinct current sheets take longer to develop (this is shown dynamically in movie 4 in the supplementary material \cite{sup}). Also, as discussed in Ref. \cite{pritchett.guidefield}, the introduction of a guide-field appears to lead to the out-of-plane current being intensified along one pair of separatrix arms. At $500.0\omega_{pe}^{-1}$ for $B_{Z0}=0.8B_{P}$ a circular area of high out-of-plane current emerges at the $X$-point. We show in section \ref{sec:vortex} that this is correlated with the emergence of a magnetic island, which, as shown in Ref. \cite{island.current}, can cause electron acceleration such that out-of-plane currents are locally increased.  

  \begin{figure}[htbp]
    \includegraphics[width=0.99\linewidth]{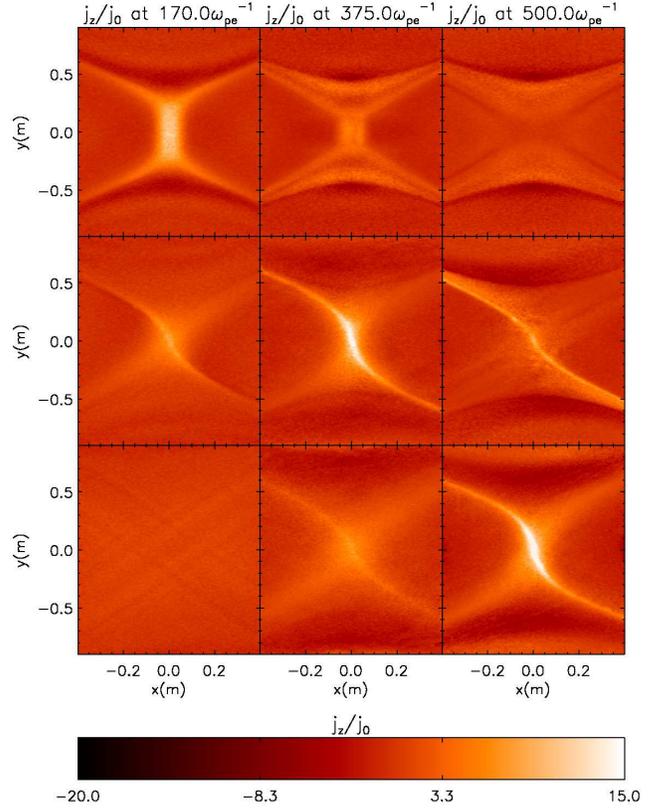}
     \caption{ \label{Fig:jzconbc}As in Fig.~\ref{Fig:jzcon} but for closed boundary conditions and with simulation times as stated.} 
      \end{figure}

Fig.~\ref{Fig:jzconbc} shows the out-of-plane current in the simulation domain for the closed boundary case for the same guide-field strengths, for snapshot times as indicated. While we observe similar results in terms of the strengthening of the current along one pair of separatrix arms and delayed onset-times, there is no indication of spiralling or clustering at the centre of the domain (this is shown dynamically in movie 5 in the supplementary material \cite{sup}). Other tests also confirmed that there was no evidence for the formation of vortices within the simulation time. However, for greater strengths of guide-field, the current sheets appeared to be thinning, which is in line with several other reconnection experiments (see Ref. \cite{birn} page 96-97). 

  \begin{figure}[htbp]
    \includegraphics[width=0.99\linewidth]{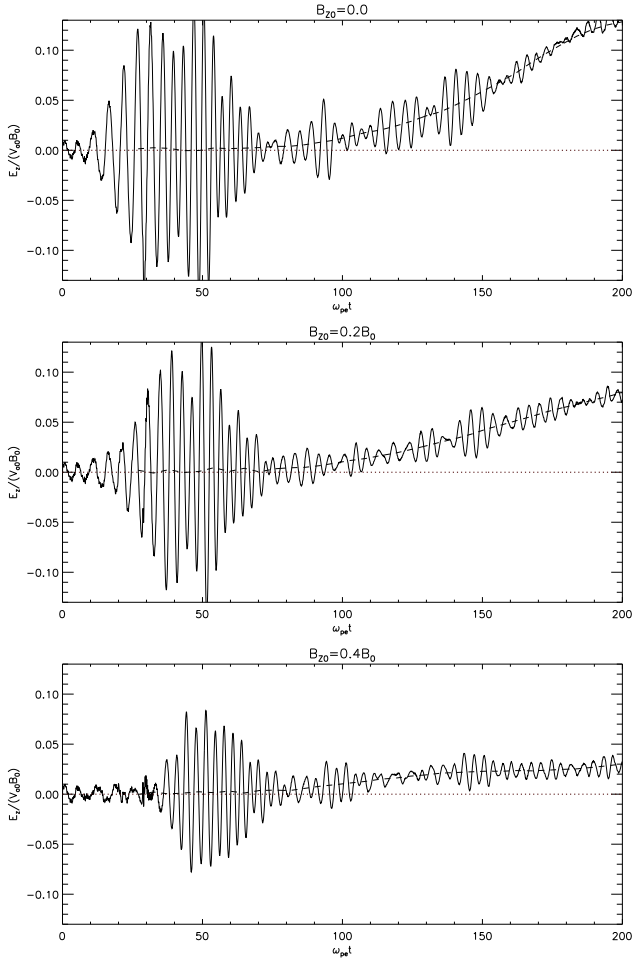}
     \caption{ \label{Fig:oscildelay} Out-of-plane electric field at the $X$-point for guide-fields of $0.0$, $0.2$ and $0.4B_P$ at the beginning of the simulation in the open boundary case. The dashed line represents the same values, smoothed with a filter width of $30\omega_{pe}^{-1}$, indicating the increase in reconnection electric field following reconnection onset.} 
      \end{figure}

Further, the introduction of guide-field had a notable effect on the oscillations in the out of-plane-electric field preceding reconnection onset. Fig.~\ref{Fig:oscildelay} shows the reconnection electric field for the open boundary case for increasing guide-field strengths, showing a smoothed line to indicate the mean increase of the field. As shown, for greater guide-field, the formation of a wave burst in the initial oscillation is increasingly delayed and the amplitudes reached are increasingly reduced. However, in each case, reconnection onset coincides with the decline of the amplitude of the initial burst. A connection appears to exist between the two events, indicative of the transient nature of reconnection in $X$-point collapse.

  \subsubsection{Vortex generation for high guide-field cases}
\label{sec:vortex}

\begin{figure}
    \includegraphics[width=0.99\linewidth]{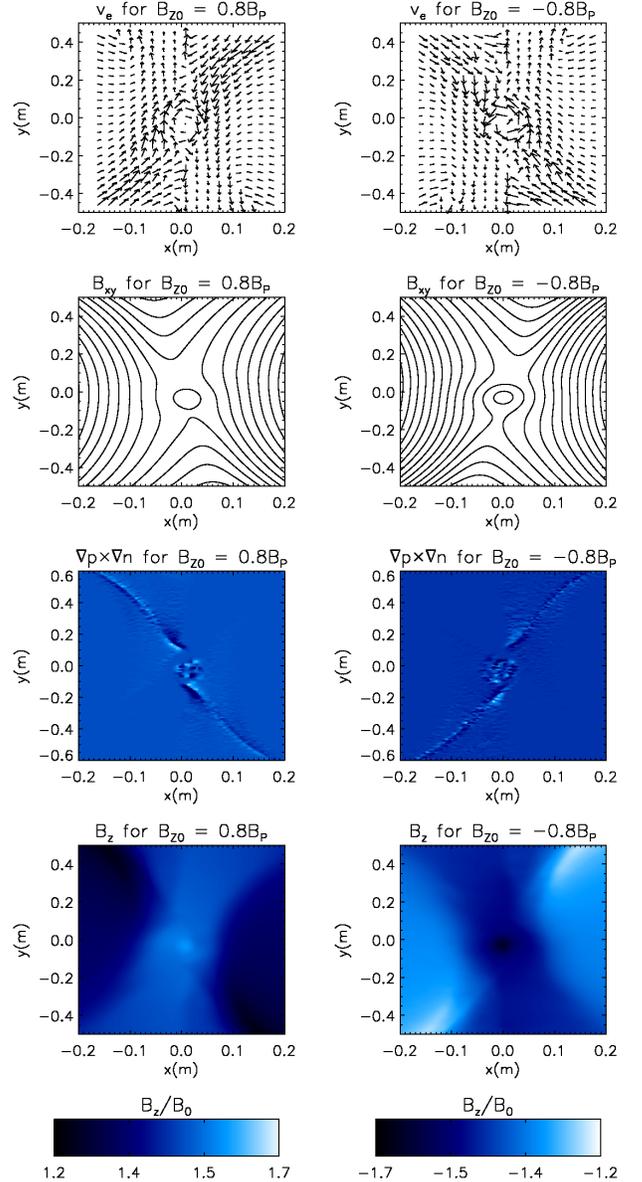}
     \caption{ \label{Fig:vortex}From top to bottom: in-plane electron velocity, in-plane magnetic field lines (magnetic flux function), the value of the $\nabla n \times \nabla p$ term and out-of-plane magnetic field in the vicinity of the vortex for guide-fields of $B_{Z0}=0.8B_{P}$ and $B_{Z0}=-0.8B_{P}$, at $500\omega_{pe}^{-1}$.} 
      \end{figure}

As pointed out, the anomalous peak in the reconnection electric field and current for a guide-field of $B_{Z0}=0.6B_{P}$, as shown in Fig.~\ref{Fig:picezjz}, is due to the emergence of vortical motion in the plane of the simulation domain. However, while in this case, the emerging vortex moved along the positive $y$ direction of the domain and dissipated, for the case of $B_{Z0}=0.8B_{P}$, the vortex remained more stable and more defined. Fig.~\ref{Fig:vortex} shows a series of plots relating to the electron vortex that emerges for a guide-field of $B_{Z0}=0.8B_{P}$ and that of a case where the guide-field is reversed, i.e. $B_{Z0}=-0.8B_{P}$, which provides a clue regarding the nature of the vortex. Further, movie 6 in the supplementary material \cite{sup} shows the change in the electron velocity grid for $B_{Z0}=0.8B_{P}$, showing the formation of the vortex (note that the film extends further than the end of simulation time used in the previous plots).

\begin{figure}
    \includegraphics[width=0.99\linewidth]{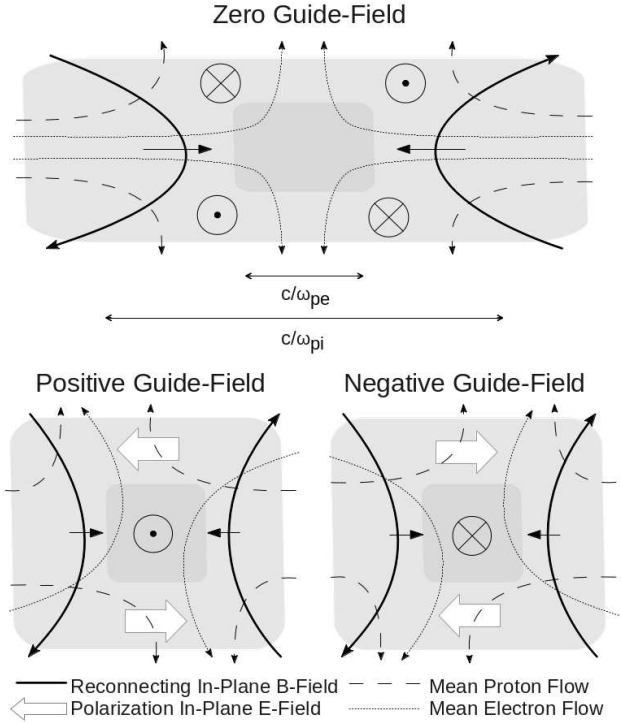}
     \caption{ \label{Fig:decouple}(Top) Adopted from \cite{birn}, chapter 3.1.1, showing a schematic of dissipation region for anti-parallel reconnection, indicating the decoupled motion of electrons and protons and how they lead to the quadrupolar out-of-plane magnetic field structure. Lengths of the ion and electron dissipation regions are indicated. (Bottom) Proposed version of the above schematic when modified by a positive or negative out-of-plane magnetic guide-field. As shown in each case, shear electron flows are induced along the field lines (see Ref. \cite{birn}, page 96-101), leading to a polarisation electric field. This in turn leads to polarisation drift of the protons across the diffusion region.} 
      \end{figure}

 Following \cite{birn}, page 89-92, the quadrupolar structure of the out-of-plane magnetic field that emerges during reconnection of opposing field lines is caused by the decoupled motion of electrons and ions in the dissipation region. This is illustrated in the top panel of Fig.~\ref{Fig:decouple}. Ions (here protons) decouple and cease to move with the field lines sooner than the electrons, leading to current loops which in turn lead to magnetic structures. In a similar fashion, it is argued here that imposing a sufficiently strong guide-field in the diffusion region would impose current loops, which in turn lead to vortical motion of particles. 
 
 As discussed in Ref. \cite{birn}, page 96-101, and shown in Ref. \cite{sato97} and Ref. \cite{pritchett.guidefield}, the presence of a guide-field in reconnection can lead to a shear flow of electrons along the in-plane magnetic field at the separatrices. This in turn leads to an ion polarization drift across the dissipation region to compensate for the resulting charge imbalance. This was also observed in the results of this study, as demonstrated in the bottom panel of Fig.~\ref{Fig:decouple}. Reversing the guide-field was shown to reverse the motion of electron flow and thus the polarization drift of the protons. Comparing this to the flows in Fig.~\ref{Fig:vortex} it is clear that the electron flows already undergo the vertical part of the vortex motion. Adding to this the influence of a substantial guide-field, and thus a $j\times B$ force, it is plausible that the guide-field, which is also present at the $X$-point, becomes a dominant factor in the decoupled motion of the electrons and initiates vortical flows as observed.   

 An additional consideration is that, as is known for shear flow of plasmas, the Kelvin-Helmholtz instability could play a role in the emergence of the vortex. In Ref. \cite{fermo} a case is made that current sheets, elongated by the influence of a guide-field, that undergo secondary reconnection are in fact triggered by the Kelvin-Helmholtz instability, and that the emergence of the magnetic islands is due to the emergence of electron vortices. Again, there are significant differences in the simulation set-up used in this study, but it is consistent in the aspect that the vortex observed in this simulation was accompanied by the formation of a magnetic island (see Fig.~\ref{Fig:vortex}). Further, movie 6 in the supplementary materials \cite{sup} shows what could be interpreted as secondary vortices, forming after the initial vortex emerged, as would be consistent with Kelvin-Helmholtz vortices. 

It is further noted that the temperature, and thus the pressure at the $X$-point also greatly increases for greater guide-fields. It could thus be argued that electromotive force could play a role in the generation of these vortices, similarly to the simulation in Ref. \cite{exb}. Since the electromotive force is generated by a $\nabla n \times \nabla p$ term, where $n$ and $p$ are number density and scalar pressure respectively, this term was plotted for the simulation domain (see Fig.~\ref{Fig:vortex}). As can be seen, there is a correlation between areas of high vorticity and the magnitude of the $\nabla n \times \nabla p$ term.

 \subsection{Initial Oscillation in Reconnection Field}
\label{sec:oscil}

\begin{figure}[htbp]
    \includegraphics[width=0.99\linewidth]{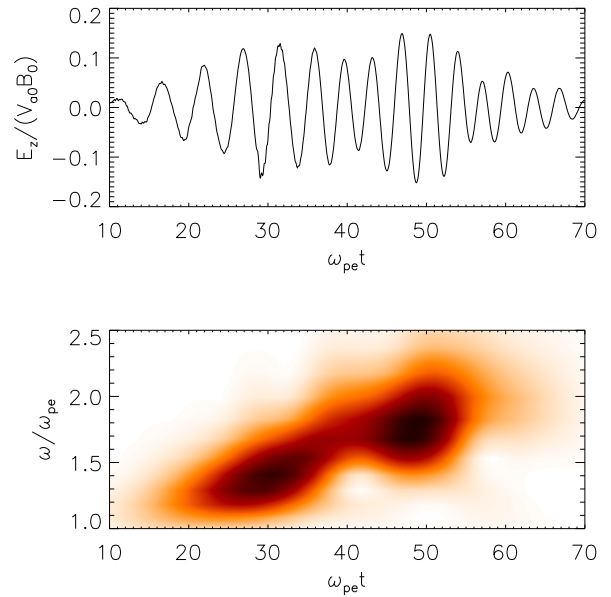}
     \caption{ \label{Fig:wavelet}(Top) The oscillation in $E_z$ at the $X$-point at the beginning of the simulation. (Bottom) The very same segment of the $E_z$ oscillation, decomposed using a wavelet function, showing the change in frequency over the same time period. The colour intensity in the plot relates to the amplitude of the oscillation. } 
      \end{figure}

Both in the open and the closed boundary case, as well as for increasing guide-fields (see Fig.~\ref{Fig:oscildelay}), the simulation exhibited a burst of rapid oscillations in the reconnection field, directly preceding reconnection onset. This oscillation was shown to be part of a more general wave structure in $E_z$, spreading over the simulation domain. While this oscillation is present in all cases of guide-field, it appears most dominant in the guide-field-free case and was thus investigated for this case. By decomposing the oscillation at the $X$-point using a wavelet function, the change in frequency was shown to rise from approximately $1\omega_{pe}$ to $2\omega_{pe}$, as shown in Fig.~\ref{Fig:wavelet}.

\begin{figure}[htbp]
    \includegraphics[width=0.99\linewidth]{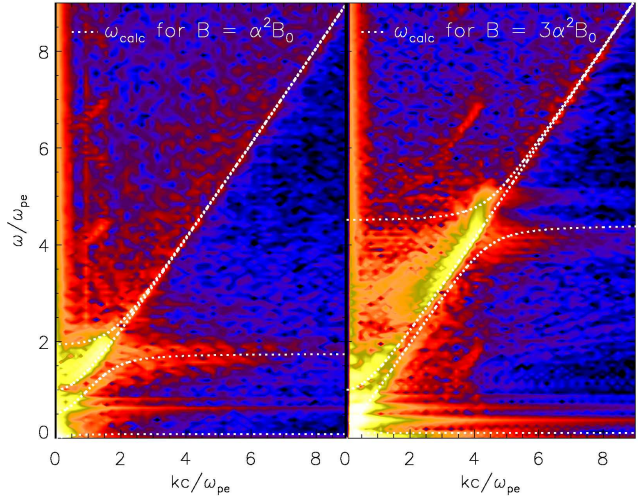}
     \caption{ \label{Fig:dispersion}(left) Dispersion plot generated by sampling $E_z$ over horizontal slice at $y = 0$ over the period in which the initial oscillations occurred, i.e. $100\omega_{pe}^{-1}$, and applying a 2D Fourier transform. (right) Same plot for a three-fold increased in-plane field. Both plots show the logarithm of the respective amplitude and have been superimposed with a simulated dispersion plot for an ideal Maxwellian plasma with identical parameters to the simulation and magnetic field strengths as indicated.} 
      \end{figure}

By taking a horizontal slice through the simulation domain at $y = 0$ and mapping $E_z$ at these locations over time, contours of wave patterns were composed showing the variation of $E_z$ with time and position. These were transformed using a 2D Fourier transform to give a plot representing the dispersion relationships of the waves. To test the consistency of this method, similar tests were carried out for simulation runs with in-plane fields of greater values of $B_0$, as in ($\ref{eqn:Bxy}$). Resonances seemed to appear, scaling linearly with the field strength (see Fig.~\ref{Fig:dispersion}).

The dispersion relationships in Fig.~\ref{Fig:dispersion} were superimposed with the standard dispersion curves for waves in a cold Maxwellian plasma, propagating perpendicularly to a magnetic field, given in Ref. \cite{dendy}, page 45, as 
    \begin{equation}
    c^2k^2 = \omega^2-\omega_{pe}^2,
    \label{eqn:ko}
    \end{equation}
i.e. the normal mode, and
    \begin{equation}
    c^2k^2 = \frac{(\omega^2-\omega_1^2)(\omega^2-\omega_2^2)}{\omega^2-\omega_{uh}^2},
    \label{eqn:kx}
    \end{equation}
i.e. the extraordinary mode. $\omega_{uh}$ represents the upper hybrid frequency and $\omega_1$ and $\omega_2$ are given respectively as
 
    \begin{equation}
    \omega_1=(\omega_{uh}^2-3\omega_{ce}^2/4)^{1/2}-\omega_{ce}/2
    \label{eqn:w1}
    \end{equation}
and
    \begin{equation}
    \omega_2 = \frac{\omega_{uh}}{2}\left\{(1+3\omega_{pe}^2/\omega_{uh}^2)^{1/2}+(1-\omega_{pe}^2/\omega_{uh}^2)^{1/2}\right\}.
    \label{eqn:w2}
    \end{equation}
The value for $\omega_{pe}$ here was set equal to that used in the simulation, while it was found that the value of $\omega_{ce}$ that best matches the results is that corresponding to a magnetic field of $\alpha^2B_0$, which corresponds to the strength of $B_y$ at the x-boundaries of the domain, as given by ($\ref{eqn:Bxy}$). This was shown to hold true for greater values of $B_0$ as shown in right panel of Fig.~\ref{Fig:dispersion}. 

From the superimposed dispersion plots in fig.~\ref{Fig:dispersion} it can be seen that the greatest wave amplitudes occur in the ordinary mode. Since these correspond to electromagnetic waves, this occurrence could have observational consequences in the radio wave regime, potentially solar radio burst fine structure spikes (see Ref. \cite{spikes1}). The frequencies of the waves range approximately from the plasma frequency up to the upper-hybrid frequency, where they reach a maximum. This relates well to \cite{spikes2}, where a model of Zebra patterns in superfine solar radio emission is proposed in which plasma oscillations generate wave emissions at the upper-hybrid frequency, propagating perpendicularly to the magnetic field and  polarised in the ordinary or extraordinary mode. Observations and further models for the generation of Zebra patterns in superfine solar radio emission, based on plasma mechanisms and also at the upper-hybrid frequency (or multiples thereof), are discussed in Ref. \cite{spikes3}. The same test was carried out by sampling over a vertical slice through the simulation domain at $x = 0$ and plots similar to Fig.~\ref{Fig:dispersion} were generated. However, in this scenario the greatest amplitude seemed to be located in the extraordinary mode, above $\omega_2$.  

Furthermore, it was found that the oscillations in the out-of-plane electric field at the $X$-point correspond to a reversal in the flow of reconnecting magnetic field lines i.e. oscillatory reconnection occurred (see movie 3 in supplementary material \cite{sup}). Oscillatory reconnection in $X$-point configurations is discussed in detail in Ref. \cite{oscilrecon2} and Ref. \cite{oscilrecon1}.  

\begin{figure}[htbp]
    \includegraphics[width=0.99\linewidth]{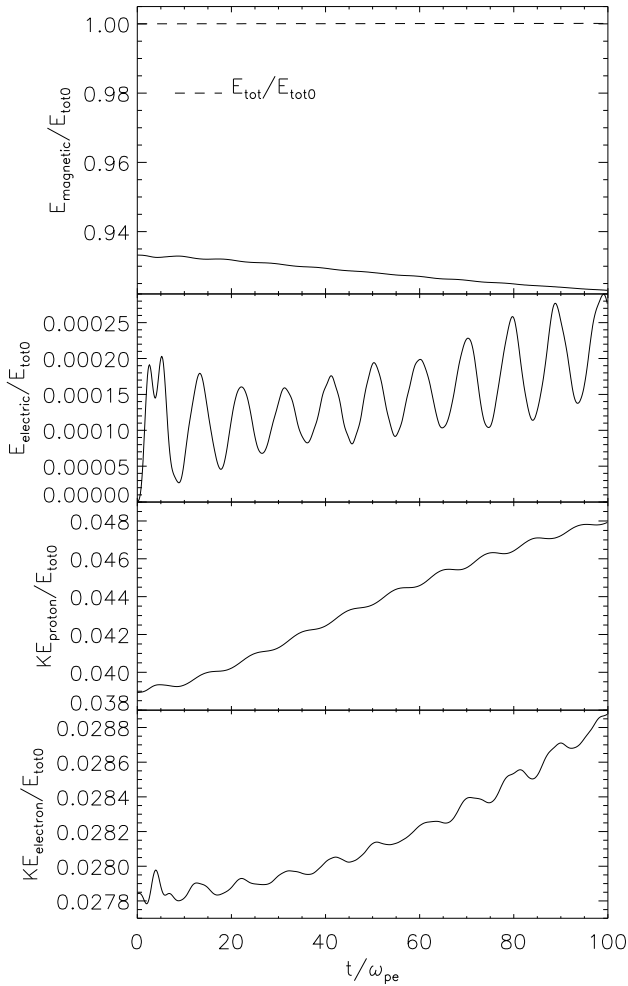}
     \caption{ \label{Fig:energies} showing (from top to bottom) the total magnetic field energy and total electric field energy, integrated over the simulation domain, and the total kinetic energies of protons and electrons, summed over all particles, all for the closed boundary case over the initial period of oscillation. Since energy is conserved in this case (see additional line in total magnetic energy plot) the increase in electric field energy (as well as kinetic energies) must be supplied by the decrease in magnetic field energy.} 
      \end{figure}

To answer from where the energy of these waves originated, the total energy of the magnetic and electric fields as well as the kinetic energies of electrons and protons was investigated in the closed boundary case (as these boundary conditions do not allow any inflows or outflows and thus conserve total energy) for the initial oscillation period, as shown in Fig. ~\ref{Fig:energies}. It is shown that the only decreasing energy component is that of the magnetic field. As the particle kinetic energies also increase it is to be concluded that the wave energy in the electric field also results from the conversion of magnetic energy. It can also be seen that kinetic energy of electrons appears to vary significantly in phase with the electric field energy. This could be linked to the role the electrons play in oscillatory reconnection. The energy conservation error, $[E(t)-E(0)]/E(0)$, was tested for these first $100\omega_{pe}^{-1}$ and was found to be $0.0001$. For the whole simulation time, i.e. $500\omega_{pe}^{-1}$, this was found to be $0.01$.
    
\section{Summary and Conclusion}
\label{sec:conclusion}

Simulations of magnetic reconnection during collisionless, stressed $X$-point collapse by D. Tsiklauri and T. Haruki \cite{Tsiklauri2007,tsiklauri2008} were extended by the inclusion of the out-of-plane magnetic guide field and the use of different boundary conditions. A kinetic, 2.5D, fully electromagnetic, relativistic particle-in-cell numerical code was used.

Two cases of boundary conditions for this simulation were defined and investigated i.e. closed boundary conditions, where particles are reflected from the boundary and magnetic flux is confined, and open boundary conditions, where particles and magnetic flux can escape through boundary. In the zero guide-field case (see Fig.~\ref{Fig:picbccompare}) it was found that for closed boundary conditions the reconnection electric field at the $X$-point reached an initial peak at approximately $170\omega_{pe}^{-1}$ and a second peak at approximately $340\omega_{pe}^{-1}$, thus resulting in a very distinct two peak profile. In the open boundary case reconnection onset was slower and a distinct peak in the reconnection field was not reached before $200\omega_{pe}^{-1}$. However, the reconnection rate was also slower to saturate, declining by less than half before the end of the simulation i.e. at $500\omega_{pe}^{-1}$. Further, the peak value of reconnection current at the $X$-point for open boundary conditions was approximately twice the value in the closed boundary case. In both cases a peak in the particle number density is observed in the proximity of initial peak values. 

Guide field values were chosen ranging from $0.2$ to $1.0$ of the in-plane magnetic field, as defined in ($\ref{eqn:Bz}$). In the closed boundary case (see Fig.~\ref{Fig:picezjzbc}) peak reconnection rates were reached sooner for greater guide-fields, however peak values were greatly reduced. In the open boundary case (see Fig.~\ref{Fig:picezjz}) the onset of reconnection was increasingly delayed for greater guide-field, however, reconnection peak rates were shown to be higher for guide-fields up to $0.8$ the in-plane field. Reductions in amplitude are also observed in the MRX, as shown in Ref. \cite{tharp2}, where this was shown to be due to a reduction in the Hall-current. Delays in on-set times observed are consistent with results in Ref. \cite{sato97} and may be indicative of more limited meandering motion of electrons. Initial increases in the reconnection current, similar to results in Ref. \cite{pritchett.guidefield}, where studied at low values of guide-field. It was possible to establish an optimal guide-field value for which the initial rate of increase of the current in the open boundary case was maximised, found to be between 0.1 and 0.2 the in-plane field. 

 An unusually high peak in the out-of-plane current at the $X$-point was observed in the open-case for a guide-field of $0.6$ the value of the in-plane field. Further it is shown in Fig.~\ref{Fig:jzcon} that, towards the end of the simulation for a guide-field of $0.8$ the in-plane field, a strong localised current forms at the $X$-point. Both of these observations are linked to the emergence of an electron vortex alongside a magnetic island at the $X$-point. It is shown in Ref. \cite{island.current} that magnetic islands forming at the $X$-point lead to electron acceleration such that a localised out-of-plane current is produced, consistent with these observations.
 
The electron vortex that emerged, for guide-fields equal or greater than $0.6$ the in-plane field, was shown to reverse vorticity if a an oppositely pointing guide-field was applied (see Fig.~\ref{Fig:vortex}). It was concluded that this could be explained by consideration of the decoupled motion of electrons and protons in the diffusion region (see Ref. \cite{birn}, chapter 3.1.1). While in the zero guide-field case quadrupolar magnetic fields emerge at the reconnection site due to current loops in the particle flows, imposing a sufficiently strong guide-field not only causes an electron shear flow along the current sheet, but the guide-field appears to impose current loops and thus vortical motion is induced. Also, it was shown that magnetic islands in the in-plane magnetic field emerge at the same location as the vortex. This coincides with the results found by \cite{fermo}, which suggest that magnetic islands in reconnection with a guide-field generally emerge as a result of vortical motion of electrons, triggered by a Kelvin-Helmholtz instability. It was also investigated whether the electro-motive force, as described in Ref. \cite{exb}, was likely to play a part in the vortex generation. It was shown that the terms responsible for the generation of electro-motive force, i.e. $\nabla n \times \nabla p$, were of greater magnitudes in the simulation domain at greater values of vorticity. This will be further investigated elsewhere. 

The initial oscillations in the out-of-plane electric field were investigated and were shown to be part of a larger wave pattern. By making cuts over the x and y axes of the simulation domain and sampling over time, contours of the wave patterns were obtained and analysed using a 2D Fourier transform (see Fig.~\ref{Fig:dispersion}). From this it was possible to conclude that the emerging waves are predominantly in the ordinary mode, in the horizontal direction, and the in the extraordinary mode, in the vertical direction, and thus could be of observational relevance to solar radio burst fine structure spikes. Waves in the ordinary mode were shown to reach frequencies of up to the upper-hybrid frequency, which corresponds well with the model for the generation of Zebra patterns in superfine solar radio emission described in Ref. \cite{spikes2}. The oscillations were also shown to be linked with oscillatory reconnection through the X-point. Further, by plotting the total energy contained in the magnetic and electric field as well as the total kinetic energy of the protons and electrons over the initial period of the oscillation in the closed (i.e. energy conserving) boundary case, it was shown that the energy of the waves must be supplied by the conversion of magnetic field energy (see Fig.~\ref{Fig:energies}).   

We would like to stress that these electric field oscillations are not an artefact of idealised boundary conditions. In the solar atmosphere, the reconnecting magnetic field is anchored to the photosphere, while being able to reconnect and change its configuration in the corona and the chromosphere. In our model, in the closed boundary case, a similar configuration is presented since constant magnetic flux at the boundary mimics partial anchoring of field lines in the photosphere, while $X$-point collapse occurs as in the corona or the chromosphere. Moreover, the oscillation also occur in the open boundary case, which is relevant to the coronal heights.

\acknowledgements
  Authors acknowledge use of Particle-In-Cell code EPOCH and support by development team (http://ccpforge.cse.rl.ac.uk/gf/project/epoch/). Computational facilities used are that of Astronomy Unit, Queen Mary University of London and STFC-funded UKMHD consortium at St. Andrews and Warwick Universities. JGVDP acknowledges support from STFC PhD studentship. DT is financially supported by STFC consolidated Grant ST/J001546/1, The Leverhulme Trust Research Project Grant RPG-311 and HEFCE-funded South East Physics Network (SEPNET).


\end{document}